\documentclass[%
 aip,
 amsmath,amssymb,
reprint,%
]{revtex4-1}

\usepackage{graphicx}% Include figure files
\usepackage{dcolumn}% Align table columns on decimal point
\usepackage{bm}% bold math
\usepackage[utf8]{inputenc}
\usepackage[LGR,T1]{fontenc}
\usepackage{mathptmx}
\usepackage{etoolbox}
\usepackage{braket}
\usepackage[frozencache=true,cachedir=minted-cache]{minted}
\usepackage{todonotes}
\usepackage{hyperref}
\usetikzlibrary{shapes.geometric}
\usetikzlibrary{decorations.pathmorphing,patterns, decorations.pathreplacing, decorations.shapes, arrows.meta}
\usepackage{textcomp}
\usepackage{lmodern, newunicodechar}
\usepackage{xcolor} % to access the named colour LightGray
\definecolor{LightGray}{gray}{0.9}
\usepackage{bbm}
\definecolor{Burgundy}{RGB}{144,0,32}

\newcommand{\e}{\mathrm{e}}
\renewcommand{\i}{\mathrm{i}}

\newcommand{\kb}{k_{B}}

\newcommand{\id}{\hat{\mathbbm{1}}}

\renewcommand{\d}{\mathrm{d}}

\newcommand{\h}{\hat{H}}
\newcommand{\ph}{\hat{h}}
\newcommand{\hint}{\hat{H}_{\text{int}}}
\renewcommand{\a}{\hat{a}}
\newcommand{\ad}{\hat{a}^{\dagger}}
\newcommand{\aw}{\hat{a}_\omega}
\newcommand{\awd}{\hat{a}_\omega^{\dagger}}
\newcommand{\bn}{\hat{b}_n}
\newcommand{\bnd}{\hat{b}_n^{\dagger}}
\newcommand{\sx}{\hat{\sigma}^x}
\newcommand{\sy}{\hat{\sigma}^y}
\newcommand{\sz}{\hat{\sigma}^z}
\newcommand{\hc}{\text{h.c.}}

\renewcommand{\eqref}[1]{Eq.~(\ref{#1})}

\newunicodechar{α}{$\mathtt{\alpha}$}
\newunicodechar{β}{$\mathtt{\beta}$}
\newunicodechar{ω}{$\mathtt{\omega}$}
\newunicodechar{∆}{$\mathtt{\Delta}$}
\newunicodechar{ψ}{$\mathtt{\psi}$}
\newunicodechar{ε}{$\mathtt{\varepsilon}$}
\newunicodechar{δ}{$\mathtt{\delta}$}
\newunicodechar{λ}{$\mathtt{\lambda}$}

\makeatletter
\def\@email#1#2{%
 \endgroup
 \patchcmd{\titleblock@produce}
  {\frontmatter@RRAPformat}
  {\frontmatter@RRAPformat{\produce@RRAP{*#1\href{mailto:#2}{#2}}}\frontmatter@RRAPformat}
  {}{}
}%
\makeatother
\begin{document}

\title[MPSDynamics.jl]{MPSDynamics.jl: Tensor network simulations for finite-temperature (non-Markovian) open quantum system dynamics}
% Force line breaks with \\
\author{Thibaut Lacroix}
 \email{thibaut.lacroix@uni-ulm.de}
 \affiliation{Institut f\"ur Theoretische Physik und IQST, Albert-Einstein-Allee 11, Universit\"at Ulm, D-89081 Ulm, Germany}

\author{Brieuc Le D\'e}%
 \email{lede@insp.jussieu.fr}
\affiliation{Sorbonne Universit\'{e}, CNRS, Institut des NanoSciences de Paris, 4 place Jussieu, 75005 Paris, France}%

\author{Angela Riva}%
 \email{angela.riva@inria.fr}
\affiliation{LPENS, Département de physique, \'Ecole normale supérieure, Centre Automatique et Systèmes (CAS), MINES ParisTech, Université PSL, Sorbonne Université, CNRS, Inria, 75005 Paris, France}%

\author{Angus J. Dunnett}%
 \email{angus.dunnett@multiversecomputing.com}
\affiliation{Multiverse Computing, 7 rue de la Croix Martre, 91120 Palaiseau, France}%

\author{Alex W. Chin}%
 \email{alex.chin@insp.jussieu.fr}
\affiliation{Sorbonne Universit\'{e}, CNRS, Institut des NanoSciences de Paris, 4 place Jussieu, 75005 Paris, France}%

\date{\today}% It is always \today, today,
             %  but any date may be explicitly specified

\begin{abstract}
The \texttt{MPSDynamics.jl} package provides an easy to use interface for performing open quantum systems simulations at zero and finite temperatures.
The package has been developed with the aim of studying \mbox{non-Markovian} open system dynamics using the state-of-the-art numerically exact Thermalized-Time Evolving Density operator with Orthonormal Polynomials Algorithm (T-TEDOPA) based on environment chain mapping.
The simulations rely on a tensor network representation of the quantum states as matrix product states (MPS) and tree tensor network (TTN) states.
Written in the \texttt{Julia} programming language, \texttt{MPSDynamics.jl} is a versatile open-source package providing a choice of several variants of the Time-Dependent Variational Principle (TDVP) method for time evolution (including novel bond-adaptive one-site algorithms).
The package also provides strong support for the measurement of single and multi-site observables, as well as the storing and logging of data, which makes it a useful tool for the study of many-body physics. 
It currently handles long-range interactions, time-dependent Hamiltonians, multiple environments, bosonic and fermionic environments, and joint system-environment observables.
\end{abstract}

\maketitle

\section{Introduction \label{sec:intro}}

The description of quantum systems as interacting with an uncontrolled external environment is central in quantum physics \cite{breuer_theory_2007, rivas_open_2011, weiss_quantum_2012, blum_density_2012}, quantum chemistry \cite{tully_perspective_2012, agostini_different_2019}, and quantum biology \cite{mohseni_quantum_2014}.
In numerous settings there is a need to go beyond the approximation of a weak coupling between the system and its environment, and beyond the Markovian approximation (i.e. memoryless environment) \cite{li_concepts_2018}.
In these cases analytical calculations are typically not available, and one has to rely on numerical methods.
However, direct many-body calculations are not tractable because of the exponential growth of the underlying Hilbert space dimension with the size of joint \{System + Environment\}. Moreover, methods based on reduced density techniques also become computationally expensive as the memory time and complexity (rank) of the system-environment interactions increase. 
In response, several numerical methods to simulate open quantum system (OQS) dynamics in this non-perturbative regime have been developed, such as the Time Evolving Matrix Product Operator (TEMPO)~\cite{strathearn_efficient_2018, strathearn_modelling_2020} method (implemented in the OQuPy package \cite{oqupy_2020}), the Automated Compression of Environments (ACE) algorithm~\cite{cygorek_simulation_2022, cygorek_sublinear_2024, cygorek_mcygorekace_2024}, the Hierarchical Equations Of Motions (HEOM) technique \cite{tanimura_numerically_2020,mangaud2023survey} (implemented in the QuTip package \cite{johansson_qutip_2013}), the Multi-Layer--Multi-Configuration Time-Dependent Hartree (ML-MCTDH) method \cite{meyer_studying_2012} (implemented in the Quantics package \cite{WORTH2020107040, quantics}), collision models \cite{ciccarello_quantum_2022}, the Dissipation-Assisted Matrix Product Factorization (DAMPF) method~\cite{somoza_dissipation-assisted_2019} and the \mbox{T-TEDOPA} technique~\cite{prior_efficient_2010, woods_simulating_2015, oviedo2016phase,schroder2016simulating, schroder2019tensor,tamascelli_efficient_2019} to name a few.
TEMPO, ACE, HEOM, and T-TEDOPA belong to the specific class of numerically exact methods (i.e. methods for which the numerical error can be estimated exactly in terms of the convergence parameters).
To our knowledge, \texttt{MPSDynamics.jl} is currently the only maintained package available implementing the T-TEDOPA method in Julia. 

In the T-TEDOPA method, the curse of dimensionality preventing a direct resolution of the Schrödinger equation for the joint \{System + Environment\} state is circumvented using a tensor network representation of the joint quantum state\cite{orus_practical_2014, bridgeman_hand-waving_2017, evenbly_tensorsnet}.
The particularity of \mbox{T-TEDOPA} is to treat the system and its environment on an equal footing, thus leaving the internal dynamics of the environment open to inspection \cite{schroder2016simulating,del2018tensor,riva_thermal_2023}.
The accuracy of this representation is controlled by the size of the bond dimension $D$ of the tensor network, and the number of environmental modes that need to be taken into account is determined by the total simulation time.

\texttt{MPSDynamics.jl} has already been used in several scientific publications on quantum thermodynamics \cite{dunnett_matrix_2021}, electron transfer \cite{dunnett_simulating_2021}, the calculation of linear absorption spectra \cite{dunnett_influence_2021,Hunter2024}, information backflow \cite{lacroix_unveiling_2021,lacroix_non-markovian_2024}, dynamical equilibrium states and the formation of polarons \cite{riva_thermal_2023}, and nanoscale dissipation \cite{lacroix_non-markovian_2024}.
Moreover the T-TEDOPA framework is currently used by several research groups to work on projects in quantum photo-chemistry \cite{lede_ESIPT_2024}, multi-dimensional non-linear optical spectra~\cite{lorenzoni_systematic_2024} or exact simulations of pigment protein complexes \cite{caycedo-soler_exact_2022}. Previous applications of the TEDOPA formalism have also been used to look at ground state properties of strongly interacting system-environment properties, where this approach has strong links to the well-established MPS implementations of DMRG and NRG \cite{chin2011generalized, blunden2017anatomy}.

We also note here that the many body T-TEDOPA approach to open quantum systems has also recently been implemented on noisy intermediate-scale quantum computers \cite{nishi2024simulation,guimaraes2023noise,guimaraes2024digital}.

The \texttt{MPSDynamics.jl} package is distributed under the GPL-3.0 free software license and is publicly available on Github at \href{https://github.com/shareloqs/MPSDynamics}{https://github.com/shareloqs/MPSDynamics}.
A Zenodo record is also publicly available~\cite{dunnett_angusdunnettmpsdynamics_2021}.\\

In Sec.~\ref{sec:overview} we give an overview of the package.
In Sec.~\ref{sec:tedopa} we summarize the important concepts of the \mbox{T-TEDOPA} method.
Section~\ref{sec:examples} presents short and simple examples using the package.
The instruction to download and install \texttt{MPSDynamics.jl} are provided in Sec.~\ref{sec:download}.
Future developments of the package are discussed in Sec.~\ref{sec:conclusion}.

\section{Package overview \label{sec:overview}}
In this section we give an overview of the abilities and usage of the package.
For a more detailed explanation, a documentation is provided with the package at \href{https://shareloqs.github.io/MPSDynamics/}{https://shareloqs.github.io/MPSDynamics/}, and several example scripts are available in the package (see on Github). For a detailed presentation of the underlying theory of T-TEDOPA in the formalism and context used in \texttt{MPSDynamics.jl}, we refer the reader to the thesis of A.~Dunnett \cite{dunnett2021tensor}. 

\subsection{States and Hamiltonians}
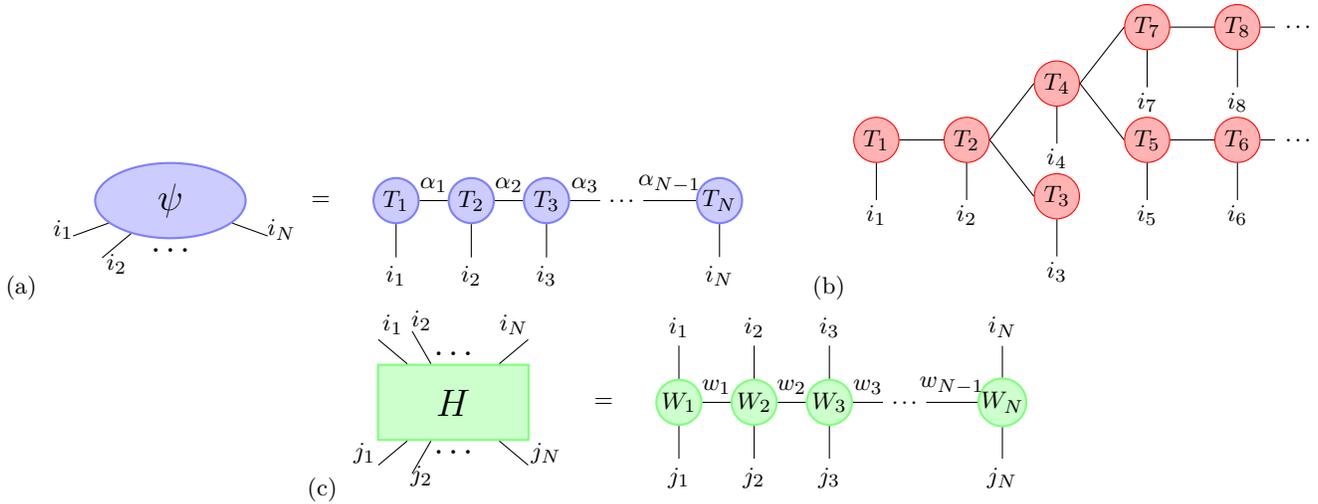
\begin{figure*}
    \centering
    (a)
    \begin{tikzpicture}
        [tensor/.style={circle,draw=blue!50,fill=blue!20,thick,
                     inner sep=0pt,minimum size=6mm}]
         \node (c) at (-2,0) [ellipse,
                        draw=blue!50,
                        fill=blue!20,
                        thick,
                        minimum width = 2cm,
                        minimum height = 1cm] {\Large $\psi$};
        \draw (c.200) -- node[label=left:$i_1$] {} +(200:0.5);
        \draw (c.220) -- node[below] {$i_2$} +(220:0.5);
        \node[below] at (c.south) {\Large $\ldots$};
        \draw (c.-20) -- node[label=right:$i_N$] {} +(-20:0.5);
        \node (eq) at (0,0) {$=$} ;
        \node[tensor] (T1) at (1,0) {$T_1$} ;
        \node (i1) at (1, -1) {$i_1$} ;
        \node[tensor] (T2) at (2,0) {$T_2$} ;
        \node (i2) at (2, -1) {$i_2$} ;
        \node[tensor] (T3) at (3,0) {$T_3$} ;
        \node (i3) at (3, -1) {$i_3$} ;
        \node (dots) at (4,0) {$\ldots$} ;
        \node[tensor] (TN) at (5.3,0) {$T_N$} ;
        \node (iN) at (5.3, -1) {$i_N$} ;
        \draw (i1) -- (T1.south) ; \draw (T1.east) -- node[above] {$\alpha_1$} (T2.west) ; 
        \draw (i2) -- (T2.south) ; \draw (T2.east) -- node[above] {$\alpha_2$} (T3.west) ;
        \draw (i3) -- (T3.south) ; \draw (T3.east) -- node[above] {$\alpha_3$} (dots.west) ;
        \draw (iN) -- (TN.south) ; \draw (dots.east) -- node[above] {$\alpha_{N-1}$} (TN.west) ;
\end{tikzpicture}
    \hspace{20px}(b)
    \begin{tikzpicture}
  [grow=right,
   level distance=15mm,
   sibling distance=8mm,
   every node/.style={draw=red,fill=red!30,circle,inner sep=1pt,minimum size=6mm},
   edge from parent path={(\tikzparentnode.east) -- (\tikzchildnode.west)}]
\node (root) {$T_1$} % Root node
      child[grow=right, level distance=12mm, sibling distance=15mm] {{node (H2) {$T_2$} % First branch
        child {node (C1) {$T_3$}}
        child {node (C2) {$T_4$}
                child {node (C3) {$T_5$}
                child {node (C4) {$T_{6}$}}
                                }
                child {node (C5) {$T_{7}$}
                child {node (C6) {$T_{8}$}}
                }
                }
      }
    };
\draw (C6) -- ++(0.5, 0) node[right, draw=none, fill=none]  {$\ldots$};
\draw (C4) -- ++(0.5, 0) node[right, draw=none, fill=none]  {$\ldots$};
\draw[-] (root) - ++(0, -8mm) node[pos=1, yshift=-2mm,draw=none, fill=none, inner sep=1pt] {$i_1$};
\draw[-] (H2) - ++(0, -8mm) node[pos=1, yshift=-2mm,draw=none, fill=none, inner sep=1pt] {$i_2$};
\draw[-] (C1) - ++(0, -8mm) node[pos=1, yshift=-2mm,draw=none, fill=none, inner sep=1pt] {$i_3$};
\draw[-] (C2) - ++(0, -8mm) node[pos=1, yshift=-2mm,draw=none, fill=none, inner sep=1pt] {$i_4$};
\draw[-] (C3) - ++(0, -8mm) node[pos=1, yshift=-2mm,draw=none, fill=none, inner sep=1pt] {$i_5$};
\draw[-] (C4) - ++(0, -8mm) node[pos=1, yshift=-2mm, draw=none, fill=none, inner sep=1pt] {$i_6$};
\draw[-] (C5) - ++(0, -8mm) node[pos=1, yshift=-2mm, draw=none, fill=none, inner sep=1pt] {$i_7$};
\draw[-] (C6) - ++(0, -8mm) node[pos=1, yshift=-2mm, draw=none, fill=none, inner sep=1pt] {$i_8$};            
\end{tikzpicture}
    \newline (c)
    \begin{tikzpicture}
        [tensor/.style={circle,draw=green!50,fill=green!20,thick,
                     inner sep=0pt,minimum size=6mm}]
        \node (c) at (-2,0) [rectangle,
                        draw=green!50,
                        fill=green!20,
                        thick,
                        minimum width = 2cm,
                        minimum height = 1cm] {\Large $H$};
        \draw (c.140) -- node[label=above:$i_1$] {} +(140:0.5);
        \draw (c.120) -- node[label=above:$i_2$] {} +(120:0.5);
        \node[above] at (c.north) {\Large $\ldots$};
        \draw (c.40) -- node[label=above:$i_N$] {} +(40:0.5);
        \draw (c.220) -- node[label=left:$j_1$] {} +(220:0.5);
        \draw (c.240) -- node[below] {$j_2$} +(240:0.5);
        \node[below] at (c.south) {\Large $\ldots$};
        \draw (c.-40) -- node[label=right:$j_N$] {} +(-40:0.5);
        \node (eq) at (0,0) {$=$} ;
        \node[tensor] (T1) at (1,0) {$W_1$} ;
        \node (i1) at (1, +1) {$i_1$} ;
        \node (j1) at (1, -1) {$j_1$} ;
        \node[tensor] (T2) at (2,0) {$W_2$} ;
        \node (i2) at (2, 1) {$i_2$} ;
        \node (j2) at (2, -1) {$j_2$} ;
        \node[tensor] (T3) at (3,0) {$W_3$} ;
        \node (i3) at (3, 1) {$i_3$} ;
        \node (j3) at (3, -1) {$j_3$} ;
        \node (dots) at (4,0) {$\ldots$} ;
        \node[tensor] (TN) at (5.3,0) {$W_N$} ;
        \node (iN) at (5.3, 1) {$i_N$} ;
        \node (jN) at (5.3, -1) {$j_N$} ;
        \draw (i1) -- (T1.north) ;\draw (j1) -- (T1.south) ; \draw (T1.east) -- node[above] {$w_1$} (T2.west) ; 
        \draw (i2) -- (T2.north) ;\draw (j2) -- (T2.south) ; \draw (T2.east) -- node[above] {$w_2$} (T3.west) ;
        \draw (i3) -- (T3.north) ;\draw (j3) -- (T3.south) ; \draw (T3.east) -- node[above] {$w_3$} (dots.west) ;
        \draw (iN) -- (TN.north) ;\draw (jN) -- (TN.south) ; \draw (dots.east) -- node[above] {$w_{N-1}$} (TN.west) ;
\end{tikzpicture}
    \caption{(a) A multi-partite quantum state $\psi$ described by the coefficients $\psi_{i_1 i_2 \ldots i_N}$ can be written as a Matrix Product State (MPS). A MPS is a product of (at-most) rank-3 tensors ${T_n}_{i_n}^{\alpha_{n-1} \alpha_n}$ where $\mathrm{dim}(i_n) = d$ is the local Hilbert space dimension, and $\mathrm{dim}(\alpha_n) = D$ is the so called `bond dimension'. (b) Tree tensor networks (TTN) can be used to represent quantum states and Hamiltonians. TTN are especially relevant for systems interacting with more than two environments. (c) Similarly, operators (such as the Hamiltonian $H$) can be decomposed into a product of rank-4 tensors ${W_n}_{j_n i_n}^{w_{n-1} w_n}$. This decomposition is called a Matrix Product Operator (MPO).}
    \label{fig:MPS}
\end{figure*}

A simulation with \texttt{MPSDynamics.jl} requires a MPS or a TTN representing the initial wave-function and a Matrix Product Operator (MPO) (or accordingly a TTN) representing the Hamiltonian.
Figure~\ref{fig:MPS} shows diagrammatic representations of these objects.
The package contains various functions for generating MPSs and MPOs used for simulating generic models such as the spin-boson model, the independent boson model, XYZ Hamiltonians, spin chains or tight-binding models for instance; but no attempt is made to be comprehensive.
Methods are provided to construct MPS and TTN representations of standard states such as product states, Fock states, displaced states, or single-particle electronic states for instance.
For generic MPO construction, one can use the \texttt{ITensors.jl} package \cite{ITensor} and convert the resulting object into a form compatible with \texttt{MPSDynamics.jl} using built-in functions.
One can also directly construct MPO tailored for the problem at hand (see App.~\ref{app:MPO}).
The nature of the Hamiltonian can be fairly generic, for instance long-range interactions~\cite{lacroix_unveiling_2021}, time-dependent Hamiltonians, bosonic and fermionic environments, multiple environments~\cite{dunnett_simulating_2021, lacroix_unveiling_2021, dunnett_influence_2021, lacroix_non-markovian_2024} can be handled.
The fact that TTN with an arbitrary number of branches are supported means that the coupling to several environments (more than 2) at different temperatures -- or even of a different nature -- can be simulated.

For finite-temperature simulations, and experimental or \textit{ab initio} bath spectral densities (SDs), an integrated implementation of the \texttt{ORTHOPOL} package \cite{gautschi_algorithm_1994} is provided in order to generate the chain parameters needed to describe the chain-mapped environment (see Sec.~\ref{sec:tedopa}) \cite{chin_exact_2010, tamascelli_efficient_2019}.

\subsection{Time-Evolution}
The time-evolution methods currently implemented belong to the family of Time-Dependent Variational Principle (TDVP) \cite{haegeman_time-dependent_2011}.
The central point of this method, in the modern tensor-network formulation, is that instead of solving the Schrödinger equation and then truncating the MPS representation of the quantum state, one can solve the equations of motion projected into a space of restricted bond dimension\cite{haegeman_time-dependent_2011, paeckel_time-evolution_2019}. 
The major advantage of this method is that it naturally preserves the unitarity of the time evolution and conserves the energy (except in its two-site implementation).
Three variants of TDVP are implemented in the \texttt{MPSDynamics.jl} package:
\begin{itemize}
    \item one-site TDVP on TTN and MPS (TDVP1),
    \item two-site TDVP on MPS (TDVP2),
    \item a variant of one-site TDVP with adaptive update of bond-dimensions on MPS (DTDVP) \cite{dunnett_efficient_2021}.
\end{itemize}
The main advantage of the one-site TDVP (TDVP1) algorithm is that it preserves the unitarity of the MPS during the time evolution. 
Its main problem, conversely, is that the time evolution is constrained to happen on a manifold constituted by tensors of fixed bond dimension, a quantity closely related to the amount of entanglement in the MPS, and such a bond dimension has therefore to be fixed before the beginning of the time evolution. 
This strategy will necessarily be non optimal: the unavoidable growth of the bond dimensions required to describe the quantum state at later times should ideally mirror the entanglement growth induced by the time evolution. 
The two-site implementation (TDVP2) operates in a similar way to Time-Evolving Block-Decimation (TEBD), and allows for a similar dynamical growth of the bond dimensions, and therefore better describes the entanglement in the MPS. 
It suffers however of other drawbacks: first of all, a truncation error is introduced (by the means of an SVD decomposition), which entails a loss of unitarity of the time-evolved MPS. 
Moreover, TDVP2 has worse scaling properties $\mathcal{O}(d^3)$ with the size of the local dimensions $d$ of the MPS. 
This is a major issue when dealing with bosons in non-perturbative OQS problems that require large local Hilbert space dimensions $d$.  
The DTDVP algorithm combines the best features of TDVP1 and TDVP2: it preserves unitarity, it has the same scaling properties $\mathcal{O}(d^2)$ as TDVP1, and it adapts the bond dimensions to the entanglement evolution at each site and at each time-step. 
DTDVP does not suffer from a truncation error, but introduces only a projection error that can be controlled with a tolerance parameter.

\begin{figure}
    \centering
    \includegraphics[width=\columnwidth]{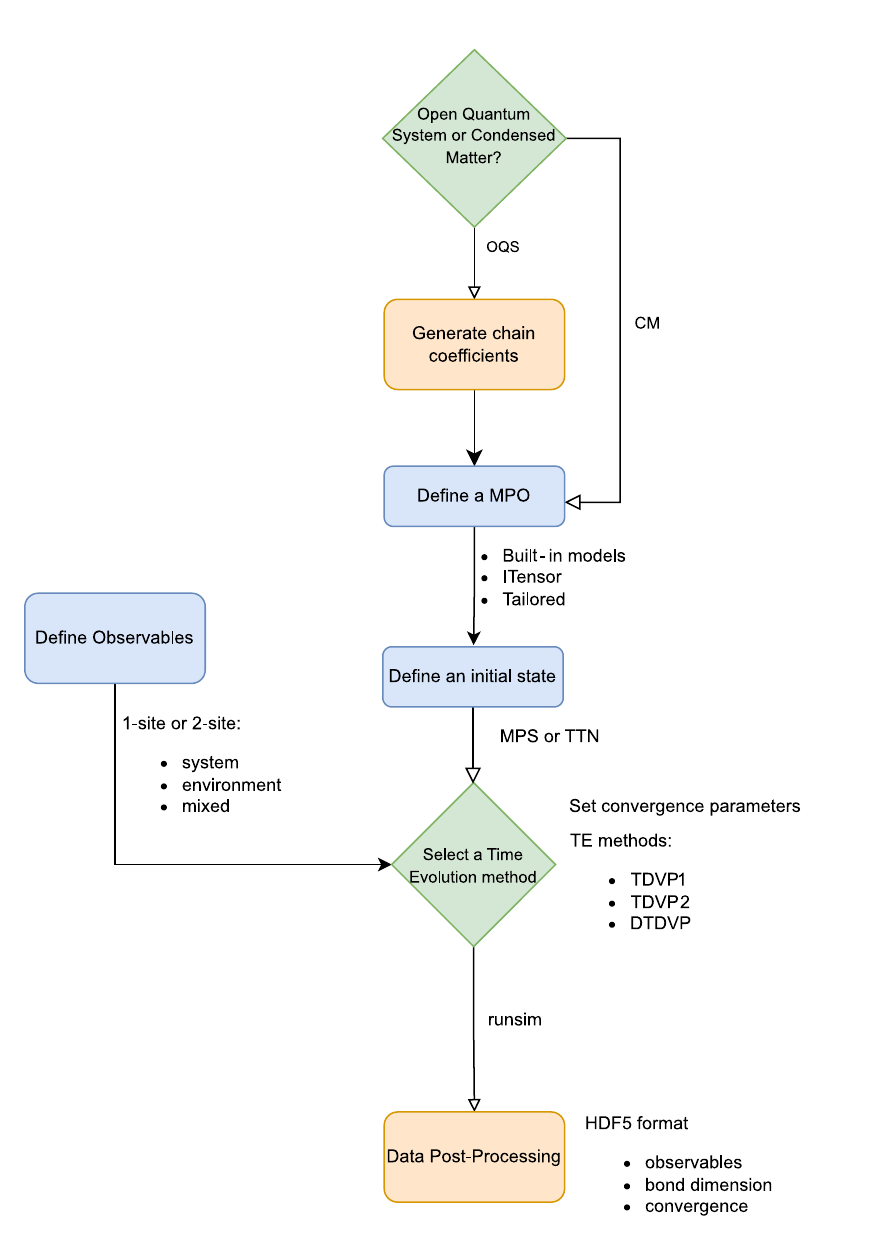}
    \caption{Package workflow diagram: the user provides a state, a Hamiltonian, observables of interest, pick a time-evolution methods, and, when the script stopped running, can analyse the dynamics of the observables.}
    \label{fig:workflow}
\end{figure}

A graphical summary of the package workflow is presented in Fig.~\ref{fig:workflow}.
We would like to highlight that, to our knowledge, \texttt{MPSDynamics.jl} is the only package implementing all three TDVP variants for MPS and TTN.
The elementary tensor operations are implemented in all cases using the \texttt{TensorOperations.jl} package \cite{TensorOperations.jl}.

\subsection{Observables and data processing}

Local one-site and two-site observables, as well as non-local two-site observables, can be efficiently computed for each time-step of the time-evolution.
System and environment observables can be computed, as well as system-and-environment non-local observables.

The data (i.e. time series of observables) is stored in the JLD format which is based on HDF5. 
Loading the data in Julia using the \texttt{JLD.jl} package will recover the full type information of the Julia variables that were stored. 
At the same time the HDF5 format is natively supported across many platforms and languages (e.g. \texttt{Python}, \texttt{Mathematica}).

\section{T-TEDOPA in a nutshell\label{sec:tedopa}}
Let us consider the following Hamiltonian where an unspecified system couples linearly  to a continuum of independent bosonic modes (normal modes)
\begin{align}
    \h &= \h_S + \int_0^{\infty} \omega\ad_\omega\a_\omega\d\omega + \hat{A}_S\int_0^{\infty}\sqrt{J(\omega)}\left(\a_\omega + \ad_\omega\right)\d\omega \label{eq:vibronic}
\end{align}
where we consider the natural units such that $\hbar = 1$.
We can introduce a unitary transformation of the continuous normal modes $\aw$ to an infinite discrete set of interacting modes $\bn$~\cite{chin_exact_2010}
\begin{align}
    \aw &= \sum_{n=0}^{\infty} U_n(\omega)\bn = \sum_{n=0}^{\infty} \sqrt{J(\omega)}P_n(\omega)\bn\ , \label{eq:chain-mapping}
\end{align}
where $P_n(\omega)$ are orthonormal polynomials such that
\begin{align}
    \int_{0}^{\infty}P_n(\omega)P_m(\omega)J(\omega)\d\omega = \delta_{n,m}\ .
\end{align}

Note that the orthonormality of the polynomials ensures the unitarity of the transformation defined in Eq.~(\ref{eq:chain-mapping}).
The mapping from a continuous set of modes to a (still infinite) discrete set might seem counter-intuitive, however it is a direct consequence of the separability of the underlying Hilbert space.
Under this transformation, the Hamiltonian in Eq.~(\ref{eq:vibronic}) becomes
\begin{align}
    \h = \h_S &+ \sum_{n=0}^{\infty}\varepsilon_n\bnd\bn + t_n(\hat{b}_{n+1}^\dagger\bn + \hc) + c_0\hat{A}_S(\hat{b}_0 + \hat{b}_0^\dagger)\ . \label{eq:chain-Hamiltonian}
\end{align}
Hence, this mapping transforms the normal bath Hamiltonian into a tight-binding Hamiltonian with on-site energies $\varepsilon_n$ and hopping energies $t_n$.
Another important consequence of this mapping is that now the system only interacts with the first mode $n=0$ of the chain-mapped environment.
The chain coefficients $\varepsilon_n$, $t_n$, and the coupling $c_0$ depend solely on the SD (see Ref.~\cite{chin_exact_2010}).
This makes chain mapping a tool of choice for describing systems coupled to environment with highly structured SD (e.g. experimentally measured or calculated \textit{ab initio})~\cite{chin_role_2013, alvertis_non-equilibrium_2019, dunnett_influence_2021, caycedo-soler_exact_2022, lorenzoni_systematic_2024}.
In this new representation, the Hamiltonian in Eq.~(\ref{eq:chain-Hamiltonian}) has naturally a 1D chain topology.
This makes its representation as a MPO and the representation of the joint \{System + Environment\} wave-function as a MPS suited~\cite{orus_practical_2014, paeckel_time-evolution_2019}.
The orthogonal polynomial-based chain mapping and the subsequent representation of the joint wave-function as a MPS (and the operators as MPO) are the building blocks of the T-TEDOPA method, one of the state-of-the-art numerically exact method to simulate the dynamics of open quantum systems especially in the non-Markovian, non-perturbative regimes both at zero and finite temperatures~\cite{prior_efficient_2010, woods_simulating_2015, tamascelli_efficient_2019, dunnett_simulating_2021, lacroix_unveiling_2021}.
A peculiarity of this method is that the finite temperature case is handled with an effective zero-temperature \{System + Environment\} but with a temperature-dependent SD~\cite{tamascelli_efficient_2019}.
This `trick' enables the wave-function description to be kept at finite-temperature at the moderate cost of increasing the number of environmental modes, with the maximal extension being a doubling of the number of modes for very high temperatures.

\section{Examples\label{sec:examples}}
We briefly present a few examples of applications to showcase the package workflow and illustrate its possibilities.
Further example scripts can be found on Github, and for each example there is a commented tutorial in the online documentation.
The following examples correspond to the current release (version 1.1) of the package.
Although examples are here given in dimensionless arbitrary units, all units are supported as long as the user checks the consistency of the script. However, explicit units as a Type such as provided by \texttt{Unitful.jl} are not currently supported.\\

In all the examples we first load the \texttt{MPSDynamics.jl} package:

\begin{minted}[tabsize=2,breaklines, bgcolor=LightGray]{matlab}
using MPSDynamics
\end{minted}
\subsection{Pure dephasing\label{sec:pure-deph}}
Let us consider the independent boson model Hamiltonian
\begin{align}
    \h = &\frac{\hat{\sigma}_z}{2}\left(\Delta E + \int_0^{+\infty}\sqrt{J(\omega)}\left(\a_\omega + \ad_\omega\right)\d\omega\right)\nonumber\\
    &+ \int_0^{+\infty}  \omega\ad_\omega\a_\omega\d\omega \label{eq:ibm}
\end{align}
where $\Delta E$ is the energy gap of a two-level system (TLS), $\hat{\sigma}_z$ is the third Pauli matrix, $\awd$ creates a bosonic excitation at frequency $\omega$.
We consider the case of an Ohmic SD $J(\omega) = 2\alpha\omega_c \left(\omega/\omega_c\right)^s H(\omega_c - \omega)$ where $\alpha$ is a dimensionless coupling strength, $s$ is the so called Ohmicity, $H(\omega)$ is the Heaviside step function, and $\omega_c$ is the bath cut-off frequency.
The system initial state is a superposition of spin-$\uparrow$ and spin-$\downarrow$ states $\ket{\psi} = \frac{1}{\sqrt{2}}\left(\ket{\uparrow} + \ket{\downarrow} \right)$, and the environment is in a Gibbs state at the inverse temperature $\beta = (\kb T)^{-1}$.
    
In general, parameters can be stored in an input text file (Path/input.txt) that can be read within a script with the command
\begin{minted}[tabsize=2,breaklines, bgcolor=LightGray]{python}
include("$Path/input.txt")
\end{minted}
\\
We first compute the chain coefficients of the environment (truncated to $N$ modes) that we will use to define the chain-mapped version of the Hamiltonian in \eqref{eq:ibm}.
\begin{minted}[tabsize=2,breaklines, bgcolor=LightGray]{python}
# chain parameters, i.e. on-site energies e_i, hopping energies t_i, and system-chain coupling c_0
cpars = chaincoeffs_finiteT(N, β; α=α, s=s)
\end{minted}
At zero temperature analytical expressions of the coefficients can be calculated \cite{woods_simulating_2015}, and are given by the function
\begin{minted}[tabsize=2,breaklines, bgcolor=LightGray]{python}
cpars = chaincoeffs_ohmic(N, α, s)
\end{minted}
Pre-computed chain coefficients can be read with the \texttt{readchaincoeffs} function.
\\

Numerically, one of the convergence parameter for simulating the chain is its number of modes. In order to evaluate the parameter $N$ to prevent unphysical reflection of excitations at the end of the chain, the user can start with a large chain and call the method
\begin{minted}[tabsize=2,breaklines, bgcolor=LightGray]{python}
findchainlength(tfinal, cpars)
\end{minted}
in order to check the converged number of mode.
A rule-of-thumb estimate based on the universal asymptotic speed of propagation of excitations on the chain~\cite{chin_exact_2010, de_vega_how_2015} is that \mbox{$N \simeq \omega_c t_\text{final}/4$} for zero-temperature environment, and \mbox{$N \simeq \omega_c t_\text{final}/2$} for finite-temperature environments.
We use these coefficients and the local dimensionality $d$ of the $N$ bosonic modes to set up the MPO representation of the Hamiltonian using a built-in function
\begin{minted}[tabsize=2,breaklines, bgcolor=LightGray]{python}
H = puredephasingmpo(∆E, d, N, cpars)
\end{minted}

The initial state is a product state of the system state $\ket{\psi}$ and the vacuum state for N environmental modes
\begin{minted}[tabsize=2,breaklines, bgcolor=LightGray]{python}
# Initial TLS in a superposition of up and down
ψ = zeros(2)
ψ[1] = 1/sqrt(2)
ψ[2] = 1/sqrt(2)

# MPS representation of |ψ>|Vacuum>
A = productstatemps(physdims(H), state=[ψ, fill(unitcol(1,d), N)...]) 
\end{minted}

We define the observables of interest
\begin{minted}[tabsize=2,breaklines, bgcolor=LightGray]{python}
ob1 = OneSiteObservable("sx", sx, 1)
ob2 = OneSiteObservable("sy", sy, 1)
ob3 = OneSiteObservable("sz", sz, 1)
\end{minted}
and we choose a bond dimension $D$ and a time-evolution method.

\begin{minted}[tabsize=2,breaklines, bgcolor=LightGray]{python}
method = :TDVP1 # time-evolution method

D = 2 # MPS bond dimension
\end{minted}

We are now all set to run the simulation
\begin{minted}[tabsize=2,breaklines, bgcolor=LightGray]{python}
A, dat = runsim(dt, tfinal, A, H;
            name = "pure dephasing model at temperature β = $(β)",
            method = method,
            obs = [ob1, ob2, ob3],
            convobs = [ob1],
            params = @LogParams(∆E, N, d, α, s, β),
            convparams = D,
            reduceddensity = true,
            verbose = false,
            save = true,
            plot = true,
            );
\end{minted}

The data is stored in dictionaries
\begin{minted}[tabsize=2,breaklines]{bash}
julia> dat["data"]
Dict{String, Any} with 4 entries:
  "sx"       => [1.0, 0.912818, 0.741759, 0.605797, 0.528792, 0.492497, 0.47976…
  "sz"       => [0.0, 0.0871825, 0.25824, 0.394201, 0.471207, 0.507503, 0.52023…
  "sy"       => [0.0, -0.0133489, -0.0588887, -0.0858181, -0.0759996, -0.048539…
  "times"    => [0.0, 0.0005, 0.001, 0.0015, 0.002, 0.0025, 0.003, 0.0035, 0.00…

\end{minted}

Figure~\ref{fig:examples}~(a) shows the simulation results for zero and finite temperatures compared with the analytical results for a MPS of fixed bond dimension. Taking advantage of the wave-function framework of our MPS method we can also access environmental degrees of freedom, for instance Fig.~\ref{fig:examples}~(b) shows the environmental correlations $\langle\ad_\omega\ad_{\omega'}\rangle(t) - \langle\ad_\omega\rangle(t)\langle\ad_{\omega'}\rangle(t)$. Local and non-local bath observables can be easily probed within the different examples. 

\subsection{Time-dependent Hamiltonian\label{sec:time-dep}}

Time-dependent Hamiltonians can be seamlessly used in simulation.
Let us consider a two-level system driven by an external field
\begin{align}
     \h_S(t) + \hint &= \frac{\omega_0}{2} \hat{\sigma}_z + \left(\Delta + \epsilon \sin(\omega_\text{drive} t) \right)\hat{\sigma}_x \nonumber\\
     & + \hat{\sigma}_x \int_0^{+\infty}\sqrt{J(\omega)}\left(\a_\omega + \ad_\omega\right)\d\omega \ .
\end{align}
One just starts by defining the \emph{time-independent} part of the Hamiltonian
\begin{minted}[tabsize=2,breaklines, bgcolor=LightGray]{python}
H = spinbosonmpo(ω0, ∆, d, N, cpars) # MPO representation of the Hamiltonian
\end{minted}
and creates separately list containing the \emph{time-dependent} Hamiltonian at the different time-steps of the simulation
\begin{minted}[tabsize=2,breaklines, bgcolor=LightGray]{python}
Ht = [ε*sx*sin(ωdrive*tstep) for tstep in timelist] # Time-dependent Hamiltonian term
\end{minted}
Then the simulation is run specifying which site of the MPO is time dependent
\begin{minted}[tabsize=2,breaklines, bgcolor=LightGray]{python}
A, dat = runsim(dt, tfinal, A, H;
            name = "Driving field on ohmic spin boson model",
            method = method,
            obs = [ob1],
            convobs = [ob1],
            params = @LogParams(N, d, α, ∆, ω0, s),
            convparams = D,
            timedep = true, # the Hamiltonian is time dependent
            Ndrive = 1, # the first site of the MPS/MPO (i.e. the system) is concerned
            Htime = Ht, # list of time-dependent terms
            verbose = false,
            save = true,
            plot = true,
            );
\end{minted}
Example simulation results obtained with this script are presented in Fig.~\ref{fig:examples}~(c) showing the non-trivial dynamics of $\hat{\sigma}_z$ and the growth of the bond dimension when using DTDVP.

\subsection{Long-range system-chain couplings}
If one considers an interaction Hamiltonian with space-dependent coupling coefficients such as 
\begin{align}
    \hint &= \sum_r \hat{A}_r \int_{-\infty}^{+\infty}\sqrt{J(\omega_k)}\left(\a_k \e^{\i k r} + \ad_k \e^{-\i k r} \right)\d k\ ,
\end{align}
the chain-mapped Hamiltonian becomes long-ranged and describes a so called \emph{correlated} environment~\cite{lacroix_unveiling_2021}
\begin{align}
    \hint & = \sum_{r, n} \hat{A}_r \left( \gamma_n(r) \bn  + \gamma_n(r)^* \bnd \right)\ ,
\end{align}
where the new system-chain coupling coefficients are given by $\gamma_n(r) = \int_{\mathbb{R}^+} J(\omega_k)\e^{\i k r} P_n(k) \d k$.
This generic type of coupling is handled through a built-in method generating the environment MPO providing a list of system component positions $R$
\begin{minted}[tabsize=2,breaklines, bgcolor=LightGray]{python}
Hchain = correlatedenvironmentmpo(R, N, d, chainparams=cpars, fnamecc="couplingcoeffs.csv", s=s, α=α)
\end{minted}
The coupling coefficients $\gamma_n(r)$ can be provided by the user. 
If not, they will be computed and stored in order to be reused in the future.

\subsection{Fermionic environments}
In addition to bosonic environments, the methods implemented in the code also allow to deal with fermionic ones.
More specifically, it is possible to implement the MPO corresponding to the a fermionic non-interacting resonance-level model \cite{jauho1994time}, which is essentially a single impurity Anderson model \cite{anderson_localized_1961}.
We give here a brief introduction of the problem in the spinless case, more information and the full commented code can be found in the online documentation.
We consider a Hamiltonian representing a local impurity ($\hat H_\text{loc}$), conduction electrons ($\hat H_\text{cond}$), where the electrons are considered to be spinless, and a hybridization term between the impurity and the conduction electrons ($\hat H_\text{hyb}$)
    \begin{align}
        \h  &= \overbrace{\epsilon_d \hat d^\dagger \hat d}^{\hat H_\text{loc}} + \underbrace{\sum_{k} V_k \Big( \hat d^\dagger \hat c_k + \hat c_k^\dagger \hat d \Big)}_{\hat H_\text{hyb}} + \underbrace{\sum_k \epsilon_k \hat c_k^\dagger \hat c_k}_{\hat H_\text{cond}}\ .
    \end{align}
    This Hamiltonian can be expressed in terms of filled and empty modes of the conduction electrons, thus effectively coupling the local defect to two environments on which a chain mapping can be performed \cite{kohn_efficient_2021}.
For two leads of $N$ electrons each, coupled to the local impurity through a spectral density function $V$, characterised by a dispersion relation $\epsilon$, the Hamiltonian MPO is initialised by first computing with a built-in function the chain coefficients, for both the filled and empty lead, and then by constructing the corresponding MPO, with another built-in function
\begin{minted}[tabsize=2,breaklines, bgcolor=LightGray]{python}
chainparams1 = chaincoeffs_fermionic(N, β, 1.0; ε, V, save=false) # empty lead (labelled "1.0")
chainparams2 = chaincoeffs_fermionic(N, β, 2.0; ε, V, save=false) # filled lead (labelled "2.0")
H = tightbinding_mpo(N, εd, chainparams1, chainparams2)
\end{minted}
Note that the chain coefficients for the two leads are computed with a thermofield-like \cite{devega_thermofield_2015} transformation, which allows to perform finite temperature simulations \cite{kohn_efficient_2021} at inverse temperature $\beta$.

\subsection{Non-adiabatic dynamics on continuous energy landscapes \label{sec:proton}}
The MPS formalism can also be used for physical chemistry problems with energy surfaces described along a coordinate \cite{lede_ESIPT_2024}. For instance, a proton transfer model is implemented with the description in space of a two-level system. The introduction of a reaction coordinate (RC) allows to recover the double-well viewpoint as well as the diabatic and adiabatic basis.
The Hamiltonian reads
\begin{align}
H_\text{S} = \omega^0_{e} |e\rangle \langle e| + \omega^0_{k} |k\rangle \langle k| + \Delta (|e\rangle \langle k| + |k\rangle \langle e|)
\end{align}
where $e$ labels an enol state and $k$ a keto state.
The RC tensor is described by
\begin{align}
\nonumber H_\text{RC} + H_\text{int}^\text{S-RC} = & \, \omega_\text{RC} (d^{\dagger}d + \frac{1}{2}) \\
&+ g_{e} |e\rangle \langle e|( d + d^{\dagger})+ g_{k} |k \rangle \langle k|( d + d^{\dagger})\ .
\end{align}
The coefficients $g$ are set up in order to recover the doublewell formalism. Built-in methods allow to couple environments to this extended system such as the MPO

\begin{minted}[tabsize=2,breaklines, bgcolor=LightGray]{python}
H = protontransfermpo(ω0e, ω0k, x0e, x0k, ∆, dFockRC, d, N, cpars, cparsRC, λreorg)
\end{minted}
ruled by the Hamiltonian
\begin{align}
\nonumber H_\text{B} + H_\text{int}^\text{RC-B} = &\int_{0}^{+\infty} \mathrm{d}k \omega_k b_k^\dagger b_k + \lambda_\text{reorg}(d + d^{\dagger})^2 \\
&- (d + d^{\dagger})\int_0^{+\infty} \mathrm{d}\omega \sqrt{J(\omega)}(b_\omega^\dagger+b_\omega) 
\end{align}
with the reorganization energy
$
\lambda_\text{reorg} = \int\d\omega J(\omega)/\omega$.

\begin{figure*}
    \centering
    \includegraphics[width=\textwidth]{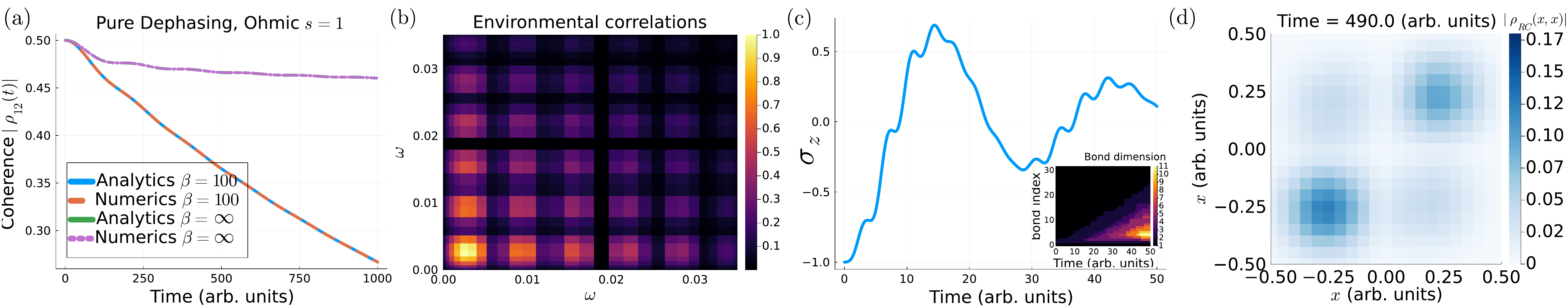}
    \caption{(a) Dynamics of the coherence in a pure dephasing model (Sec.~\ref{sec:pure-deph}) at zero and finite temperatures for the analytical results and simulations performed with a MPS of bond dimension $D = 2$. (b)  As we have access to the whole joint \{System + Bath\} state we can compute non-trivial bath observables such as the environmental correlations $\langle\ad_\omega\ad_{\omega'}\rangle(t) - \langle\ad_\omega\rangle(t)\langle\ad_{\omega'}\rangle(t)$. (c) Dynamics of the population of a dissipative TLS driven by an external time-dependent field (Sec.~\ref{sec:time-dep}). (Inset) The adaptive one-site TDVP method enables the bond dimension to be increased on the fly. (d) Reduced density matrix expressed in the reaction coordinate dimension (Sec.~\ref{sec:proton}) obtained from the MPS representing the \{electronic system + RC + bath\}. }
    \label{fig:examples}
\end{figure*}

Wave-packet dynamics and non-adiabatic transitions can easily be simulated and analyzed with the reduced density matrix. After the dynamics, the reduced density matrix can be expressed in the reaction coordinate dimension, resulting in a visualisation of the wave-packet in space, as shown in Fig.~\ref{fig:examples}~(d).

\subsection{Tree Tensor Networks}
Using TTN, several independent environments can be included.
A TTN is a generalisation of the MPS wherein each site, instead of being connected to only one other may be connected to any arbitrary number of child sites (see Fig.~\ref{fig:MPS}). 
The sites of a TTN are usually referred to as nodes, and the first node is known as the head-node.
A TTN is constructed by adding branches to specified nodes.
If one considers a TLS coupled with three bosonic environments, as has been done, for instance, to model the linear absorption spectrum of Methylene blue~\cite{dunnett_influence_2021} and the absorption and emission spectra of Proflavin in a variety of different solvents \cite{Hunter2024}, the TTN can easily be constructed as follows
\begin{minted}[tabsize=2,breaklines, bgcolor=LightGray]{python}
# Built-in TTN Hamiltonian with 3 baths
H = methylbluempo(e1, e2, δ, N1, N2, N3, d1, d2, d3, cpar1, cpar2, cpar3) 

# Initial TTN state where the head-node (the system) is connected to 3 branches corresponding to 3 chain-mapped environments
A = TreeNetwork(Any[productstatemps(1, 2, D, state=:FullOccupy)])
addtree!(A,1,TreeNetwork(productstatemps(N1, d1, D, state=:Vacuum)))
addtree!(A,1,TreeNetwork(productstatemps(N2, d2, D, state=:Vacuum)))
addtree!(A,1,TreeNetwork(productstatemps(N3, d3, D, state=:Vacuum)))
\end{minted}
This MPO and MPS illustrate the Methylene blue model with the system Hamiltonian
\begin{align}
    H_\text{S} = \left( e_2 - e_1 \right) \ket{s_2}\bra{s_2} + \delta \left( \ket{s_1} \bra{s_2} + \ket{s_2} \bra{s_1} \right)
\end{align}
The system is then interacting with three different environments 
\begin{align}
 \nonumber   H_\text{int}^\text{S-1} = &\int_{0}^{+\infty} \mathrm{d}k \omega_k b_k^\dagger b_k + \left(\ket{s_1} \bra{s_1}\right)\int_0^{+\infty}\sqrt{J_1(\omega)}\left(b_\omega + b^\dagger_\omega\right)\d\omega \\
  \nonumber   H_\text{int}^\text{S-2} = &\int_{0}^{+\infty} \mathrm{d}k \omega_k d_k^\dagger d_k + \left(\ket{s_2} \bra{s_2}\right)\int_0^{+\infty}\sqrt{J_2(\omega)}\left(d_\omega + d^\dagger_\omega\right)\d\omega \\
\nonumber    H_\text{int}^\text{S-3} = &\int_{0}^{+\infty} \mathrm{d}k \omega_k a_k^\dagger a_k \\
   &+ \left(\ket{s_1} \bra{s_2} + \ket{s_2} \bra{s_1}\right)\int_0^{+\infty}\sqrt{J_3(\omega)}\left(a_\omega + a^\dagger_\omega\right)\d\omega 
\end{align}
The \texttt{runsim} method can then be called as usual.

\section{Download and Installation \label{sec:download}}
The package may be straightforwardly downloaded and installed by typing the following command into a \texttt{Julia} REPL
\begin{minted}[tabsize=2,breaklines]{bash}
julia> ]
pkg> add https://github.com/shareloqs/MPSDynamics.git
\end{minted}

\section{Conclusion\label{sec:conclusion}}
The open-source \texttt{MPSDynamics.jl} package is a user-friendly package for the zero-and-finite temperature simulation of the  dynamics of many-body systems using tensor-network techniques.
The package is primarily designed for the study of open quantum systems in the non-Markovian, non-perturbative, regime.
Nevertheless, its applicability is wider by construction (e.g. many-body physics, non-equilibrium dynamics).
Implementing the state-of-the-art numerically exact T-TEDOPA method, \texttt{MPSDynamics.jl} enables to keep the full \{System + Environment\} information, and to put under scrutiny the environmental degrees of freedom.
The package naturally handles long-range interactions, time-dependent Hamiltonians, bosonic and fermionic environments, multiple environments, and correlated initial states.
Several widely used model Hamiltonians are already implemented as built-in functions (see methods section of the different MPOs in the online documentation for a complete list), and we encourage users to push their particular models of interest to make this a live repository for timely open system problems. 
The time-evolution methods conserve the unitarity of the dynamics and are adaptive, and will also be updated to follow new theoretical advances in time-evolution methods, particularly in the case of tree tensor networks, where long-range interactions might be tackled more effectively with entanglement renormalization methods (work in progress) \cite{schroder2019tensor}. A straight forward implementation of DMRG-like methods to find interacting ground states will also allow dynamics arising from highly correlated/entangled system-environment grounds states to explored in the future, which is a major challenge for reduced density matrix approaches, due to the inherently non-Gaussian nature of the bath in such initial conditions \cite{breuer_theory_2007}.        
The package and its documentation are available on Github at \href{https://github.com/shareloqs/MPSDynamics}{https://github.com/shareloqs/MPSDynamics}, and several example scripts are provided.
Everyone interested is welcome to contribute to development of the package.
Interested readers can either contact the authors by e-mail or open an issue on Github.
\texttt{MPSDynamics.jl} is developed continuously and future releases will be devoted to its further improvement.
Future features will include integration on GPUs \cite{Hunter2024, lambertson_computing_2024}, better parallelization, automated test units, `black box' computation of continuous wave and time-resolved optical spectra from \textit{ab initio} modelling, and Markovian closures~\cite{nuseler_fingerprint_2022}.

\begin{acknowledgments}
\noindent
TL acknowledges support from the ERC Synergy grant HyperQ (Grant No. 856432) and the EU project SPINUS (Grant No. 101135699).\\
BLD acknowledges iSiM (Initiative Sciences et ing\'enierie mol\'eculaires) from the Alliance Sorbonne Universit\'e for funding.\\
AR acknowledges support from the ANR grants HAMROQS and MECAFLUX (French Research Agency) and by Plan France 2030 through the project ANR-22-PETQ-0006. \\
AWC acknowledges ANR Project RadPolimer ANR-22-CE30-0033.
\end{acknowledgments}

\appendix
\section{How to construct a tailored MPO?}
\label{app:MPO}
To construct a Matrix Product Operator one can use a `black box' approach using \texttt{ITensors.jl} and then convert the resulting object into a form compatible with \texttt{MPSDynamics.jl} using built-in function.
Alternatively, it is possible to analytically derive the MPO representation of a Hamiltonian $\h$ which is made of a sum of local terms using the recurrence relation presented in Ref.~\cite{paeckel_time-evolution_2019}.

To define the $k$\textsuperscript{th} tensor of the MPO, we have to decompose the Hamiltonian into $\h_{k-1}^{L}$ that describes what happens before the bond $k$ (which is the bond connecting site $k$ and site $k+1$), $\h_{k+1}^R$ after the bond $k$, and $\sum_a \ph^L_{k\ a}\otimes\ph^R_{k\ a}$ at bond $k$
\begin{equation}
    \label{eq:partition}
    \h = \h_{k-1}^{L}\otimes\id^R_k + \id_k^L\otimes\h_{k+1}^R + \sum_a \ph^L_{k\ a}\otimes\ph^R_{k\ a}
\end{equation}
where $\id^R_k = \underbrace{\id\otimes\ldots\otimes\id}_{N-k+1\text{~times}}$ and $\id^L_k = \underbrace{\id\otimes\ldots\otimes\id}_{k\text{~times}}$.

A recurrence relation between the right parts of the Hamiltonian at two consecutive sites can be defined:
\begin{equation}
    \label{eq:recurrence}
    \begin{pmatrix}
    \h^R_k \\ \ph^R_k\\ \id^R_k
    \end{pmatrix} =
    \underbrace{\begin{pmatrix}
    \id_{k+1} & \hat{C}_{k+1} & \hat{D}_{k+1} \\
    \textbf{0} & \hat{A}_{k+1} & \hat{B}_{k+1} \\
    \textbf{0} & \textbf{0} & \id_{k+1}
    \end{pmatrix}}_{W_{k+1}}
    \begin{pmatrix}
    \h^R_{k+1} \\ \ph^R_{k+1}\\ \id^R_{k+1}
    \end{pmatrix}
\end{equation}
where the blocks $A_k$, $\ldots$, $D_k$ are operator-valued, $\id_k$ is an identity operator on the local Hilbert space of site $k$ and $\mathbf{0}$ is a notation for tensor elements equal to zero.
The number of columns in $A_k$ and $C_k$ is equal to size of the set ran through by the index $a$ in Eq.~(\ref{eq:partition}), which also corresponds to the number of lines of $A_k$ and $B_k$.
An interpretation can be associated with the different blocks: $D_k$ is the on-site energy, $C_k$ corresponds to the local contribution of the coupling between site $k$ and sites to its right, $B_k$ corresponds to the local contribution of the coupling between site $k$ and sites to its left, and $A_k$ corresponds to long-range coupling terms~\cite{lacroix_unveiling_2021}.

The matrices $W_k$ define the MPO representation of the Hamiltonian
\begin{equation}
    \h = \sum_{\{\sigma\}\\ \{\sigma^{'}\} \\ \{w\}} W^{\sigma_1\sigma^{'}_1}_{1\ w_0w_1}\ldots  W^{\sigma_N\sigma^{'}_N}_{N\ w_{N-1}w_N} \ket{\sigma_1\ldots\sigma_N}\bra{\sigma_1'\ldots\sigma_N'} 
\end{equation}
A schematic of an MPO representation of an operator is shown in Fig.~\ref{fig:MPS}~(c).

Below are two pedagogical examples for well-known condensed matter Hamiltonians: the $XYZ$-Hamiltonian and the Hubbard model.
Further Hamiltonians and their actual implementations (including all the built-in methods mentioned in the paper) can be found in the \texttt{src/models.jl} file inside of the package.
The interested reader can also find a worked-out example for Hamiltonians with long-range two-body interactions in Ref.~\cite{lacroix_unveiling_2021}.

\subsection{The $XYZ$-Hamiltonian}

Let us consider a $XYZ$-Hamiltonian with a nearest neighbours interaction on a 1D lattice with $N$ sites with an external field $\vec{h} = (h_x, 0, h_z)$:
\begin{equation}
    \label{eq:xyzNN}
    \h = \sum_i J_x \sx_{i}\sx_{i+1} + J_y\sy_{i}\sy_{i+1} + J_z\sz_{i}\sz_{i+1} + h_x\sx_{i} + h_z\sz_{i}\ .
\end{equation}
This Hamiltonian is local and can be partitioned around a bound $k$ in the form of \eqref{eq:partition}
\begin{eqnarray}
    \h &=& \sum_{i<k}\Big(J_x \sx_{i}\sx_{i+1} + J_y\sy_{i}\sy_{i+1} + J_z\sz_{i}\sz_{i+1}\nonumber\\
    &&+ h_x\sx_{i} + h_z\sz_{i}\Big) + h_x\sx_{k} + h_z\sz_k \nonumber\\
    && + \sum_{i>k}\Big(J_x \sx_{i}\sx_{i+1} + J_y\sy_{i}\sy_{i+1} + J_z\sz_{i}\sz_{i+1} \nonumber\\
    && + h_x\sx_{i} + h_z\sz_{i}\Big) \nonumber\\
    && + J_x \sx_k\sx_{k+1} + J_y\sy_k\sy_{k+1} + J_z\sz_k\sz_{k+1}\ .
\end{eqnarray}
Hence, we can identify $\ph_{k\ a}^R = \hat{\sigma}^a_{k+1} = \hat{\sigma}^a$ acting on the right side of the considered bond.
The recurrence relation in \eqref{eq:recurrence} gives
\begin{align}
    \h_k^R = &\h_{k+1}^R + \underbrace{J_x\sx_{k}\sx_{k+1} + J_y\sy_{k}\sy_{k+1} + J_z\sz_{k}\sz_{k+1}}_{J_a\hat{\sigma}^a\otimes\hat{\sigma}^a = \hat{C}_{k+1}^a\ph^R_{k+1\ a}}\nonumber\\
    &+\underbrace{h_x\sx_{k} + h_z\sz_{k}}_{\hat{D}_{k+1}}\ ,\\
    \ph^R_{k\ a} &= \underbrace{0}_{\hat{A}^a_{k+1}\ph^R_{k+1\ a}} + \underbrace{\hat{\sigma}^a_{k+1}}_{\hat{B}^a_{k+1}}\ .
\end{align}
Hence $\hat{A} = \textbf{0}$, $\hat{B} = \left(\begin{smallmatrix}\sx \\ \sy \\ \sz \end{smallmatrix} \right)$, $\hat{C} = (J_x\sx\ J_y\sy\ J_z\sz)$ and $\hat{D} = h_x\sx + h_z\sz$.

The on-site tensor has a bond dimension $D=5$ and a physical dimension (i.e. the dimension of the local Hilbert space) $d=2$
  \begin{equation}
    W_{k} =
    \begin{pmatrix}
    \id_{2\times2} & J_x\sx & J_y\sy & J_z\sz  & h_x\sx + h_z\sz \\
    \textbf{0} & \textbf{0} & \textbf{0} & \textbf{0} & \sx\\
    \textbf{0} & \textbf{0} & \textbf{0} & \textbf{0} & \sy\\
    \textbf{0} & \textbf{0} & \textbf{0} & \textbf{0} & \sz\\
    \textbf{0} & \textbf{0} & \textbf{0} & \textbf{0} & \id_{2\times2}
    \end{pmatrix}\ .\label{eq:on-site}
\end{equation}
The first tensor $W_1$ will a `row-tensor' (i.e. a $1\times D\times 2\times 2$-tensor) equal to the first row of \eqref{eq:on-site}, and the last tensor $W_N$ will be a `column-tensor' (i.e. a $D\times 1\times 2\times 2$-tensor) equal to the last column of \eqref{eq:on-site}.
This MPO is implemented the following way
\begin{minted}[tabsize=2,breaklines, bgcolor=LightGray]{python}
N = 100 # number of spins

D = 5 # bond-dimension of the MPO

u = unitmat(2) # 2x2 identity matrix

W = zeros(ComplexF64, D, D, 2, 2) # MPO on-site tensor

W[1,1,:,:] = W[D,D,:,:] = u
W[1,D,:,:] = hx*sx + hz*sz
i = 2
W[1,i,:,:] = Jx*sx
W[i,D,:,:] = sx
i += 1
W[1,i,:,:] = Jy*sy
W[i,D,:,:] = sy
i += 1
W[1,i,:,:] = Jz*sz
W[i,D,:,:] = sz

# the Hamiltonian is a list of all the on-site tensors
H = Any[W[1,:,:,:], fill(W, N-2)..., W[:,D,:,:]]
\end{minted}

\subsection{The Hubbard model}
The Hubbard model is described by the following Hamiltonian
\begin{equation}
    \label{eq:hubbard}
    \h = -t\sum_{i,\sigma}\hat{c}^\dagger_{i\sigma}\hat{c}_{i+1\sigma} + \hc + U\sum_i \hat{n}_{i\uparrow}\hat{n}_{i\downarrow}
\end{equation}

The on-site tensor can directly be constructed by analogy with the $XYZ$-Hamiltonian by considering that we have four operators with a nearest neighbours coupling ($\hat{c}^\dagger_{k\uparrow}$, $\hat{c}^\dagger_{k\downarrow}$, $-\hat{c}_{k\uparrow}$ and $-\hat{c}_{k\downarrow}$ ) with the coupling $-t$ and an external field $U\hat{n}_{k\uparrow}\hat{n}_{k\downarrow}$.\\

The on-site tensor has a bond dimension $D=6$ and a physical dimension (dimension of the one site Hilbert space) $d=2$
  \begin{equation}
    W_{k} =
    \begin{pmatrix}
    \id_{2\times2} & -t\hat{c}^\dagger_{k\uparrow} & -t\hat{c}^\dagger_{k\downarrow} & t\hat{c}_{k\uparrow} & t\hat{c}_{k\downarrow} & U\hat{n}_{k\uparrow}\hat{n}_{k\downarrow} \\
    \textbf{0} & \textbf{0} & \textbf{0} & \textbf{0} & \textbf{0}  & \hat{c}_{k\uparrow}\\
    \textbf{0} & \textbf{0} & \textbf{0} & \textbf{0} & \textbf{0}  & \hat{c}_{k\downarrow}\\
    \textbf{0} & \textbf{0} & \textbf{0} & \textbf{0} & \textbf{0}  & \hat{c}^\dagger_{k\uparrow}\\
    \textbf{0} & \textbf{0} & \textbf{0} & \textbf{0} & \textbf{0}  & \hat{c}^\dagger_{k\downarrow}\\
    \textbf{0} & \textbf{0} & \textbf{0} & \textbf{0} & \textbf{0}  & \id_{2\times2}
    \end{pmatrix}\ .
\end{equation}

\color{black}

\end{document}